\pdfoutput=1

\documentclass[10pt,conference]{IEEEtran}

\usepackage[utf8]{inputenc}
\usepackage[T1]{fontenc}

\usepackage{amsmath,amssymb,amsfonts,amsthm,mathtools}
\usepackage{csquotes}
\usepackage{algorithmic}
\usepackage{graphicx}
\usepackage{textcomp}
\usepackage{flushend}
\usepackage[x11names,svgnames,usenames,dvipsnames,table,rgb]{xcolor}
\usepackage[textsize=tiny,loadshadowlibrary,shadow,disable]{todonotes}

\usepackage{braket}
\usepackage{arydshln}
\usepackage{multirow}
\usepackage{float}
\usepackage{color}
\usepackage{soul}

\makeatletter
\let\MYcaption\@makecaption
\makeatother
\usepackage[font=footnotesize]{subcaption}
\makeatletter
\let\@makecaption\MYcaption
\makeatother

\usepackage{tikz}
\usetikzlibrary{arrows}
\usetikzlibrary{positioning}  
\usetikzlibrary{shadows}
\usetikzlibrary{arrows,backgrounds,calc,chains,
	matrix,positioning,shapes,shapes.geometric,
	shapes.arrows, decorations.pathmorphing, decorations.pathreplacing}

\tikzstyle{vecArrow} = [thick, decoration={markings,mark=at position
	1 with {\arrow[semithick]{open triangle 60}}}]
 
\usetikzlibrary{positioning,calc,fit,backgrounds,matrix,shapes,arrows.meta,graphs,quotes,automata,shapes.geometric,chains}

\tikzset{%
	font={\footnotesize},
	vertex/.style={draw,circle,inner sep=0pt,minimum width=0.5cm,minimum height=0.5cm,font=\small, scale=0.9},
	terminal/.style={draw,fill=white,rectangle,inner sep=2pt,font=\footnotesize,very thick},
	define color/.code={\definecolor{hsb#1}{Hsb}{#1, 1, 0.75}},
	medge/.style n args={3}{
			line width={#1pt},
			define color={#2},
			draw=hsb#2,
			out=#3,
			in=90
		},
	edge/.style 2 args={
			line width={#1pt},
			define color={#2},
			draw=hsb#2
		},
	edge0/.style 2 args={
			line width={#1pt},
			define color={#2},
			draw=hsb#2,
			out=-130,
			in=90
		},
	edge1/.style 2 args={
			line width={#1pt},
			define color={#2},
			draw=hsb#2,
			out=-50,
			in=90
		},
	zerostub/.style={
			inner sep=0,
			minimum size=3pt,
			circle,
			fill=black
		},
	cross/.style={cross out, draw=black,
			minimum size=2*(#1-\pgflinewidth),
			inner sep=0pt, outer sep=0pt,rotate=45},
	cross/.default={3pt},
	edgeOne/.style={color=RedEdge,ultra thick},
	edgeOneState/.style={color=RedEdge, thick},
	edgeMOne/.style={color=BlueEdge,ultra thick},
	edgeSqrt/.style={color=RedEdge, thick},
	edgeMSqrt/.style={color=BlueEdge, thick},
	edgeFrac/.style={color=RedEdge, thin},
	edgeOver/.style={dotted, color=blue, ultra thick},
	qubit/.style={draw,circle,inner sep=0pt,minimum width=0.35cm,minimum height=0.35cm,font=\footnotesize, thin}
}
\definecolor{RedEdge}{RGB}{191,40,40}
\definecolor{BlueEdge}{RGB}{40,191,191}
\definecolor{Blue01}{rgb}{0.26,0.39,0.85}
\definecolor{Yellow12}{rgb}{1.0,0.88,0.1}
\definecolor{Gray23}{rgb}{0.66,0.66,0.66}
\definecolor{myyellow}{RGB}{255, 253, 0}
\definecolor{myorange}{RGB}{238, 135, 51}
\definecolor{myblue}{RGB}{47, 112, 137}

\usepackage[compat=0.6]{yquant}
\useyquantlanguage{groups}

\usepackage[
	backend=biber,
	style=ieee,
	sortcites=true,
	url=true,
	eprint=true,
	giveninits=false,
	minnames=1,
	maxnames=3
	]{biblatex}

\usepackage{hyperref}

\newtheorem{example}{Example}

\def\equationautorefname~#1\null{Eq.~(#1)\null}

\AtEveryBibitem{
	\clearname{editor}%
	\clearfield{series}%
	\clearfield{isbn}%
	\clearfield{issn}%
	\clearfield{day}%
	\clearfield{month}%
	\clearfield{place}%
	\clearlist{location}%
	\clearfield{eventtitle}%
	\ifentrytype{inproceedings}{%
		\clearlist{publisher}
	}{}%
	\ifentrytype{software}{%
	}{%
		\clearfield{url}%
		\clearfield{urlyear}%
	}%
}

\begin{document}

\title{Towards Hamiltonian Simulation \\ with Decision Diagrams\vspace{-.4cm}
}
	\author{
		\IEEEauthorblockN{%
			Aaron Sander\IEEEauthorrefmark{1}\hspace{1cm}%
			Lukas Burgholzer\IEEEauthorrefmark{1}\hspace{1cm}%
			Robert Wille\IEEEauthorrefmark{1}\IEEEauthorrefmark{4}%
		}
		\IEEEauthorblockA{\IEEEauthorrefmark{1}Chair for Design Automation, Technical University of Munich, Munich, Germany}
		\IEEEauthorblockA{\IEEEauthorrefmark{4}Software Competence Center Hagenberg (SCCH) GmbH, Hagenberg, Austria}
		\IEEEauthorblockA{
			\href{mailto:aaron.sander@tum.de}{aaron.sander@tum.de},
			\href{mailto:lukas.burgholzer@tum.de}{lukas.burgholzer@tum.de},
			\href{mailto:robert.wille@tum.de}{robert.wille@tum.de}}%
		\IEEEauthorblockA{\url{https://www.cda.cit.tum.de/research/quantum/}\vspace{-.4cm}}
	}


\maketitle

\begin{abstract}
This paper proposes a novel approach to Hamiltonian simulation using \emph{Decision Diagrams} (DDs), which are an exact representation based on exploiting redundancies in representations of quantum states and operations. While the simulation of Hamiltonians has been studied extensively, scaling these simulations to larger or more complex systems is often challenging and may require approximations or new simulation methods altogether. DDs offer such an alternative that has not yet been applied to Hamiltonian simulation. In this work, we investigate the behavior of DDs for this task. To this end, we review the basics of DDs such as their construction and present how the relevant operations for Hamiltonian simulation are implemented in this data structure---leading to the first \mbox{DD-based} Hamiltonian simulation approach. Based on several series of evaluations and comparisons, we then discuss insights about the performance of this complementary approach. Overall, these studies show that DDs indeed may offer a promising new \mbox{data structure} which, for certain examples, can provide orders of magnitudes of improvement compared to the state-of-the-art, yet also comes with its own, fundamentally different, limitations. 
\end{abstract}


\section{Introduction}
Hamiltonian simulation is a powerful tool for understanding the behavior of complex physical systems \cite{cirac_goals_2012, feynman_simulating_1982, georgescu_quantum_2014, daley_practical_2022} and related problems that can be mapped to Hamiltonians \cite{lucasIsingFormulationsMany2014, zhang_mapping_2023, leipold_constructing_2021}. However, the development of efficient and accurate simulation methods remains a significant challenge due the exponential growth of the number of complex values needed to represent quantum states and operators. Hence, straightforward solutions such as statevector simulators \cite{Qiskit, bergholmPennyLaneAutomaticDifferentiation2022, CirqPythonFramework} quickly run into scalability issues. Dedicated data structures such as tensor networks \cite{orus_tensor_2019, bridgeman_hand-waving_2017} or neural network quantum states \cite{carleoSolvingQuantumManybody2017, schmittQuantumManybodyDynamics2020, gutierrezRealTimeEvolution2022} remain limited in their ability to handle large and complex systems without reaching limits in memory or runtime requirements---motivating the research for further complementary methods which can overcome this limit for certain classes of problems~\cite{xu_herculean_2023, zhou_what_2020}.

In this paper, we introduce \emph{Decision Diagrams} (DDs) as a new data structure for simulating Hamiltonians. DDs have historically found great success in the domain of classical computing as an efficient means to represent and manipulate Boolean functions and, hence, classical circuits and systems~\cite{minatoZerosuppressedBDDsSet1993, van_dijk_tagged_2017}.
Inspired by their application in classical computing, DDs were adapted to the quantum realm as a core data structure for quantum circuit simulation~\cite{grurl_noise-aware_2023, zulehnerAdvancedSimulationQuantum2019},
quantum circuit synthesis~\cite{zulehnerOnepassDesignReversible2018, saeediSynthesisQuantumCircuits2011, niemannEfficientSynthesisQuantum2014},
quantum circuit verification~\cite{burgholzerAdvancedEquivalenceChecking2021}, and measurements with shallow circuits~\cite{hillmich_decision_2021}.
however, they have yet to be applied to more general quantum systems. Their main advantage rests on exploiting redundancies in quantum states and operations which can lead to significant compression in the memory requirements needed to represent these structures and reduce the runtime necessary to perform calculations. In this paper, we shed light on their ability to simulate Hamiltonians and propose them as a complementary tool to 
other state-of-the-art methods.

To this end, we first review the basics of Hamiltonian simulation in \autoref{sec:background} and describe the limits for simulating Hamiltonian systems using the current state-of-the-art methods in \autoref{sec:GeneralProblem}. Motivated by that, we then introduce DDs as a novel, complementary approach in \autoref{sec:Proposal}---aiming for a \mbox{self-contained} coverage without assuming any computer science background. Eventually, this leads to the first Hamiltonian simulation approach based on DDs.

Following this, we conducted several series of evaluations and comparisons to gain insight into the performance of this complementary approach. The obtained results (summarized in \autoref{sec:Results}) clearly show that, for certain examples, \mbox{DD-based} Hamiltonian simulation can provide orders of magnitude of improvement compared to the \mbox{state-of-the-art}, i.e., methods based on sparse calculations, statevector simulators, and tensor networks. We also investigated the limitations of this proposed alternative, which are fundamentally different for DDs compared to the current state-of-the-art methods---requiring a shift in perspective to fully utilize their benefits. 
We conclude from these investigations that DDs offer a promising new data structure for Hamiltonian simulation that is complementary to existing approaches and, hence, may continue to play an important role in the development of efficient and accurate simulation methods.

%

\section{Hamiltonian Simulation}\label{sec:background}
In this section, the main concepts of Hamiltonian simulation are reviewed.
While the individual descriptions are kept brief, the interested reader is referred to \cite{lloydUniversalQuantumSimulators1996, hatanoFindingExponentialProduct2005,childsTheoryTrotterError2021} for more information.

\subsection{Quantum Dynamics}
\label{sec:HamiltonianSimulationBackground}
\emph{Hamiltonian simulation} is a computational technique used to simulate the dynamics of physical systems. 
The energy of a system is encoded into a mathematical object called a \emph{Hamiltonian} that describes the interactions and states of a system. 
Based on this, the dynamics of a system can be determined by solving the time-dependent Schrödinger equation for some quantum state $\ket{\Psi}$ and 
Hamiltonian $H$, i.e.,
\begin{equation}
    \label{eq:Schrödinger}
    \frac{d}{dt} \ket{\Psi} = -\frac{i}{\hbar} H \ket{\Psi(t)}.
\end{equation}
This results in the unitary time evolution operator $U(t)$ that is used to calculate the state at some time $t$, i.e.,
\begin{equation}
    \label{eq:TimeEvolution}
    \ket{\Psi(t)} = e^{-i H t} \ket{\Psi(0)} \eqqcolon U(t) \ket{\Psi(0)},
\end{equation}
where $\hbar = 1$ for simplicity.
It is then possible to calculate the dynamics of measurable physical quantities called \emph{observables} using the expectation value, i.e.,
\begin{equation}
    \label{eq:ExpectationValue}
    \braket{O(t)} = \braket{\Psi(t) | O | \Psi(t)} = \braket{\Psi(0) | U^\dagger (t) O  U(t) | \Psi(0)}.
\end{equation}
Examples of physical quantities that can be analyzed from observables include the energy of a system, magnetization, and the number of particles in a system \cite{griffiths_introduction_2005}.

\subsection{Product Formula}
\label{sec:ProductFormula}
The matrix exponential defining the time evolution operator in \autoref{eq:TimeEvolution} is often difficult or impossible to compute, especially as it grows in size and complexity. 
In order to perform the time evolution within a controlled error bound, the \emph{Lie product formula} can be used to break this operator into simpler, more efficient operations \cite{lloydUniversalQuantumSimulators1996, hatanoFindingExponentialProduct2005}. 
The matrix exponential is decomposed into its constituent terms based on the \emph{Baker-Campbell-Hausdorff} (BCH) formula, i.e.,
\begin{equation}
    \label{eq:BCH}
    e^{xA}e^{xB} = e^{x(A+B)+\frac{1}{2} x^2[A, B] + \mathcal{O}(x^3)},
\end{equation}
where $[A, B]$ is the \emph{commutator} of $A$ and $B$ defined by $[A, B]\coloneqq AB - BA$.

This results in a matrix analog to the exponential identity $e^a e^b = e^{(a+b)}$ which is commonly referred to as the \mbox{first-order} \emph{Suzuki-Trotter decomposition}, i.e.,
\begin{equation}
    \label{eq:Trotter}
    e^{x(A+B)} = e^{xA} e^{xB} + \mathcal{O}(x^2).
\end{equation}
From this, it directly follows that
\begin{equation}
    \label{eq:LieProduct}
    e^{(A+B)} = \left(e^{\frac{A+B}{n}}\right)^n = \lim_{n \to \infty} \left( e^{\frac{A}{n}} e^{\frac{B}{n}} \right)^n.
\end{equation}
Thus, the unitary time evolution for a Hamiltonian described by a sum of terms $H=A+B$ can be approximated by a sequence of discrete timesteps $\delta t = \frac{t}{n}$, i.e.,
\begin{equation}\label{eq:unitaryEvol}
        U(t) = e^{-iHt} \approx \Bigl(e^{-i \delta t A} e^{-i \delta t B} \Bigr)^n = \Bigl( U(\delta t) \Bigr)^n,
\end{equation}
where $n$ is known as the \emph{Trotter number}.
According to \autoref{eq:BCH}, this approximation is exact whenever $A$ and $B$ commute since in that case $[A, B] = 0$.
In general, the approximation error (known as \emph{Trotter error}) scales with the number of non-commuting terms in the Hamiltonian $H$ (according to \autoref{eq:BCH}).
Therefore, Hamiltonians with more non-commuting terms require more Trotter steps to compensate for the error when using product formulas. 
However, this also means that product formulas perform especially well for simulating Hamiltonians with commuting or nearly-commuting terms and, in such cases, only require few Trotter steps.

For an overview on more precise error bounds that take into account more specific parameters such as system size, see \cite{childsTheoryTrotterError2021}.

\subsection{Ising Model}\label{sec:ising}
This paper will primarily focus on the \emph{transverse field Ising model} as a representative of an important class of Hamiltonians.
In its general form, it models the spins of interacting particles and is described by the Hamiltonian
\begin{equation}
    \label{eq:GeneralizedIsing}
    H = -\sum_{\langle i, j \rangle} J_{ij} \sigma_x^{[i]} \sigma_x^{[j]} - \sum_{j} g_j \sigma_z^{[j]},
\end{equation}
where $J_{ij}$ is the interaction strength between nearest-neighbor spin pairs at sites $\langle i, j \rangle$ and $g_j$ is an external field pointing perpendicular to the interactions at site $j$. 
It is a natural starting point for Hamiltonian simulation since it can be used to formulate many computationally hard problems, such as spin glasses, \emph{Quadratic Unconstrained Binary Optimization problems} (QUBOs), or graph partitioning~\cite{lucasIsingFormulationsMany2014}.
The following constrained version of this type of Hamiltonian will be used to illustrate all proposed concepts and methods throughout the remainder of this paper.

\begin{example}\label{ex:Ising}
    The $L$-site finite 1D Ising chain is defined by
    \begin{equation}
        \label{eq:IsingModel}
        H = -J \sum_{\ell=0}^{L-2} \sigma_x^{[\ell]} \sigma_x^{[\ell+1]} - g \sum_{\ell=0}^{L-1} \sigma_z^{[\ell]},
    \end{equation}
    where the parameters are site-independent, i.e., $J_{ij} \coloneqq J$ and $g_\ell \coloneqq g$.
    Using the product formula, a single Trotter under this model has the  form
    \begin{equation}\label{eq:HamiltonianCircuitDecomposition}
        U(\delta t) = \prod_{\ell=0}^{L-2} e^{i \, J \delta t \,\sigma^{[\ell]}_x \sigma^{[\ell+1]}_x} \prod_{\ell=0}^{L-1} e^{i \, J \delta t \, \sigma^{[\ell]}_z},
    \end{equation}
    where each term is defined by the rotation gates
    \begin{equation}
        \begin{split}
            R_{xx}(\theta) & = e^{-i  \frac{\theta}{2} (\sigma_x \otimes \sigma_x)} \\
            & = \begin{psmallmatrix}
            \cos(\frac{\theta}{2}) & 0 & 0 & -i\sin(\frac{\theta}{2}) \\
            0 & \cos(\frac{\theta}{2}) & -i\sin(\frac{\theta}{2}) & 0 \\
            0 & -i\sin(\frac{\theta}{2}) & \cos(\frac{\theta}{2}) & 0 \\
            -i\sin(\frac{\theta}{2}) & 0 & 0 & \cos(\frac{\theta}{2}) \\
            \end{psmallmatrix}
        \end{split}
    \end{equation}
    and
    \begin{equation}
        \begin{split}
            R_{z}(\theta) & = e^{-i \frac{\theta}{2} \sigma_z} \\
            & = \begin{psmallmatrix}
            e^{-i\frac{\theta}{2}} & 0 \\
            0 & e^{i\frac{\theta}{2}}
            \end{psmallmatrix}.
        \end{split}
    \end{equation}
    For a 4-site chain, this decomposition is equivalent to the circuit form with rotation gates applied first to even then odd sites as sketched in the following figure:
    \begin{center}
        \resizebox{0.8\linewidth}{!}{
        \begin{tikzpicture}
        \begin{yquantgroup}
            \registers {
				qubit {} q[4];
			}
            \circuit {
                box {$U(\delta t)$} (q[-]);
            }
            \equals[$\equiv$]
			\circuit{
                box {$R_{xx}(-2J \delta t)$} (q[0], q[1]);
                box {$R_{xx}(-2J \delta t)$} (q[2], q[3]);
                box {$R_{xx}(-2J \delta t)$} (q[1], q[2]);
                box {$R_{z}(-2g \delta t)$} q;
            }
        \end{yquantgroup}
    \end{tikzpicture}}
    \end{center}
\end{example}

\subsection{Heisenberg Model}
This paper also considers a related model called the Heisenberg model \cite{heisenberg_zur_1928} which considers additional spin couplings along the y- and z-axes. This is represented by the Hamiltonian
\begin{equation}
    \label{eq:GeneralizedHeisenberg}
    \begin{split}
    H = & - J_x \sum_{\langle i, j \rangle}  \sigma_x^{[i]} \sigma_x^{[j]} - J_y \sum_{\langle i, j \rangle}  \sigma_y^{[i]} \sigma_y^{[j]} \\
    & - J_z\sum_{\langle i, j \rangle}  \sigma_z^{[i]} \sigma_z^{[j]} - h \sum_{j} \sigma_z^{[j]}.
    \end{split}
\end{equation}
\begin{example}\label{ex:Heisenberg}
    The XXX Heisenberg model is a subset of Heisenberg models such that $J_x = J_y = J_z$. The $L$-site finite 1D XXX Heisenberg chain is defined by
    \begin{equation}
        \label{eq:HeisenbergModel}
        \begin{split}
        H = & -J \Bigl( \sum_{\ell=0}^{L-2} \sigma_x^{[\ell]} \sigma_x^{[\ell+1]} + \sum_{\ell=0}^{L-2} \sigma_y^{[\ell]} \sigma_y^{[\ell+1]} \\
        & + \sum_{\ell=0}^{L-2} \sigma_z^{[\ell]} \sigma_z^{[\ell+1]} \Bigr) - h \sum_{\ell=0}^{L-1} \sigma_z^{[\ell]}.
        \end{split}
    \end{equation}
    For a 4-site chain, a single Trotter step decomposition is equivalent to the circuit as shown in the following figure:
    \begin{center}
        \resizebox{\linewidth}{!}{
        \begin{tikzpicture}
        \begin{yquantgroup}
            \registers {
				qubit {} q[4];
			}
            \circuit {
                box {$U(\delta t)$} (q[-]);
            }
            \equals[$\equiv$]
			\circuit{
                box {$R_{xx}(-2J \delta t)$} (q[0], q[1]);
                box {$R_{xx}(-2J \delta t)$} (q[2], q[3]);
                box {$R_{xx}(-2J \delta t)$} (q[1], q[2]);

                box {$R_{yy}(-2J \delta t)$} (q[0], q[1]);
                box {$R_{yy}(-2J \delta t)$} (q[2], q[3]);
                box {$R_{yy}(-2J \delta t)$} (q[1], q[2]);

                box {$R_{zz}(-2J \delta t)$} (q[0], q[1]);
                box {$R_{zz}(-2J \delta t)$} (q[2], q[3]);
                box {$R_{zz}(-2J \delta t)$} (q[1], q[2]);

                box {$R_{z}(-2g \delta t)$} q;
            }
        \end{yquantgroup}
    \end{tikzpicture}}
    \end{center}
\end{example}

\section{Motivation}\label{sec:GeneralProblem}
Hamiltonian simulation, reviewed above, is a crucial tool for understanding the behavior of quantum systems under different conditions. The physics research that uses quantum simulation can help us explore new phenomena and deepen our understanding of the fundamental workings of quantum systems. At the same time, quantum simulation can also have significant implications for material science and chemistry, enabling researchers to design and develop new materials and study chemical reactions at the molecular level.

By simulating the behavior of quantum systems using Hamiltonians, researchers can gain insights into the dynamics of quantum systems, and better understand how they will behave under different conditions. This is an important step in bridging theory and experimental results and can help accelerate the development of new quantum devices and quantum hardware. By simulating the behavior of these devices, researchers can optimize their designs and improve their performance.

Another important application of Hamiltonian simulation is in the optimization of complex systems. Many problems in fields such as machine learning \cite{sherrington_neural_1993} and logistics \cite{martonak_quantum_2004} can be mapped to Hamiltonians, allowing researchers to explore the energy landscapes and time-dynamics within given constraints. This can help to identify optimal solutions to complex problems and drive innovation in fields ranging from drug design to financial modeling.

\subsection{General Problem}

However, despite this broad spectrum of applications, Hamiltonian simulation, at its core, relies on descriptions of quantum systems that grow exponentially with system size. Quantum states and operators, such as those that represent Hamiltonians, are often described by complex vectors and matrices, respectively. For example, for a system consisting of $L$ $d$-level systems, the quantum state $\ket{\Psi}$ can be represented by a vector in $\mathbb{C}^{d^L}$, and the operator $O$ can be represented by a matrix in $\mathbb{C}^{d^L \times d^L}$. As the system size and complexity of interactions increase, these objects become increasingly computationally difficult to simulate due to the exponential growth in memory requirements and calculation runtime.

Therefore, there is a need for dedicated data structures that can efficiently simulate quantum systems on classical computers while scaling well with system size and complexity. This involves developing new algorithms and techniques that can represent the exponential growth in a more manageable way. This makes it possible to study more complex phenomena and design new materials and devices. As quantum technologies continue to advance, these classical methods will become increasingly important for simulating and understanding quantum systems in a wide range of applications.

\begin{example}\label{ex:circuit}
    Consider again the scenario from \autoref{ex:Ising} and, for simplicity, let $J=g=1$.
    Assume that the system starts in the all-zero state, i.e., $\ket{\Psi(0)} = \ket{0\dots 0}$ and that we are interested in the expectation value of the observable $O=\sigma_z^{[1]}$ at $t=1$.
    Then, using a single Trotter step and, hence, \mbox{$\delta t = \frac{t}{n} = 1$}, this corresponds to the computation as shown in the following figure:
    \begin{center}
        \resizebox{0.99\linewidth}{!}{
        \begin{tikzpicture}
        \begin{yquant}
            qubit {$\ket{0}$} q[3];
            box {$R_{xx}(-2)$} (q[0], q[1]);
            box {$R_{xx}(-2)$} (q[1], q[2]);
            box {$R_{z}(-2)$} q;
            barrier (q[-]);
            z q[1];
            barrier (q[-]);
            box {$R_{z}(2)$} q;
            box {$R_{xx}(2)$} (q[1], q[2]);
            box {$R_{xx}(2)$} (q[0], q[1]);
            output {$\bra{0}$} q;
        \end{yquant}
    \end{tikzpicture}}
    \end{center}
    This can be applied for multiple values of $t$ such that we can sample various points of the time evolution of the observable. Although simple in theory, calculating the expectation value for even a single time becomes computationally expensive for large system sizes.
\end{example}

\subsection{Related Work}
\label{sec:ClassicalQuantumSimulation}
In the past, various classical methods have been proposed to tackle the underlying complexity of this problem in different fashions. Each simulation method has its own strengths and limitations, and the choice of which method to use depends on the specific requirements of the simulation such that they can be seen as complementary to each other.

More precisely, a direct method that can save memory and computational time is by utilizing sparse vectors and matrices that only store non-zero terms \cite{berry_efficient_2007}. This method does not require the decomposition of the time evolution operator into local operations, but it still suffers from exponential scaling according to system size and requires direct construction of the unitary time evolution operator. This method is easily accessible as many linear algebra packages such as SciPy have direct support for sparse data structures and operations such as the sparse matrix exponential \cite{2020SciPy-NMeth}.

Statevector simulation \cite{Qiskit, CirqPythonFramework, bergholmPennyLaneAutomaticDifferentiation2022} is another method used for simulating quantum systems by decomposing large unitaries into circuits as done in the product formula described in \autoref{sec:ProductFormula}. This method is very useful for simulating small systems and is not limited by long-range interactions or deep circuits. They do, however, grow exponentially with system size as it is still necessary to store all the amplitudes of the statevector.

Tensor networks, on the other hand, were developed to address the exponential scaling problem of statevector simulation in quantum many-body systems \cite{orus_tensor_2019, bridgeman_hand-waving_2017}. They use local tensors to describe the quantum state and operations such that their complexity grows linearly with the system size based on an Area Law \cite{hastingsAreaLaw2007}. Tensor networks can represent larger system sizes than statevector methods, but complex operations such as deep circuits or Hamiltonians with long-range interactions require larger tensors to represent, leading to operations becoming computationally expensive and requiring high memory storage \cite{vidalTN2003}. The \emph{Time-Evolving Block Decimation} (TEBD) algorithm, which is based on the product formula as described in \autoref{sec:ProductFormula}, is limited to short-range interactions, shallow circuits, and one-dimensional geometries. The optimization of the contraction order of the operations in this method is an active area of research \cite{grayHyperoptimizedTensorNetwork2021, huang_efficient_2021, haghshenasOptimization2021, schindlerAlgorithms2020, meiromOptimizing2022}. Furthermore, the values in the tensors can be truncated, leading to an approximation of the state \cite{schollwockDensitymatrixRenormalizationGroup2011}.

There are also other relevant state-of-the-art simulation methods that are not based on the product formula. For example, Krylov subspace methods, such as the \emph{Time-Dependent Variational Principle} (TDVP, \cite{haegemanTimedependentVariationalPrinciple2011}), are not limited to short-range interactions, but they are more computationally expensive. Neural network quantum states (NQS, \cite{carleoSolvingQuantumManybody2017, schmittQuantumManybodyDynamics2020, gutierrezRealTimeEvolution2022}) are also a promising method for approximate simulation using Monte Carlo methods, but they may require exponentially many samples for large system sizes \cite{linScalingNeuralNetwork2022}.

While all of these methods have pushed our simulation capabilities forward, each has its own limitations such that there is no perfect answer to all problems. In this regard, there continues to be a need for alternative methods which can expand these capabilities further.

\section{Decision Diagrams} \label{sec:Proposal}

In this section, we introduce a novel, complementary approach to classically simulate Hamiltonians based on (quantum) \emph{Decision Diagrams} (DDs).
DDs are a data structure inspired by \emph{Binary Decision Diagrams} (BDDs) commonly used to represent and manipulate Boolean functions in computer science~\cite{minatoZerosuppressedBDDsSet1993}.
More specifically, a decision diagram, as it is considered in this work, is a directed acyclic graph with complex-valued edge weights that can be used to represent and manipulate quantum states and operators.
DDs have already shown promise in classical quantum circuit simulation~\cite{grurl_noise-aware_2023, zulehnerAdvancedSimulationQuantum2019}, synthesis~\cite{zulehnerOnepassDesignReversible2018, saeediSynthesisQuantumCircuits2011, niemannEfficientSynthesisQuantum2014}, verification~\cite{burgholzerAdvancedEquivalenceChecking2021}, and measurements with shallow circuits~\cite{hillmich_decision_2021}.
However, their application to Hamiltonian simulation has not been explored.
The remainder of this section presents an introduction to decision diagrams as a data structure for Hamiltonian simulation.
Our focus is to make this introduction as self-contained as possible without assuming any computer science background.


\subsection{Representing Quantum States}

The journey towards representing quantum states as decision diagrams starts with the simple case of a single-site system.
The state $\ket{\Psi}$ of such a system is described by two \mbox{complex-valued}, normalized amplitudes $\alpha_0$ and $\alpha_1$, i.e.,
\begin{equation}\label{eq:ssstate}
	\ket{\Psi} = \alpha_0 \ket{0} + \alpha_1 \ket{1},
\end{equation}
where $\ket{0}$ and $\ket{1}$ are the computational basis states of a \mbox{two-level} system.
This is commonly represented as a statevector
\begin{equation}
	\ket{\Psi}\equiv \begin{pmatrix} \alpha_0 & \alpha_1	\end{pmatrix}^\top.
\end{equation}
A rather simple observation and consequence of \autoref{eq:ssstate} is that this vector can be equally split into a contribution of the $\ket{0}$ state ($\alpha_0$) and a contribution of the $\ket{1}$ state ($\alpha_1$), i.e., 
\begin{equation}\label{eq:splitting}
        \bigl(
            \overbrace{
            \overset{\ket{0}}{\begin{matrix} \alpha_0
            \end{matrix}}
            \ \ \
            \overset{\ket{1}}{\begin{matrix} \alpha_1
            \end{matrix}}
            }^{\ket{\Psi}}
        \bigr)^\top.
\end{equation}
This decomposition lies at the core of the decision diagram formalism.
The decision diagram representing $\ket{\Psi}$ has the structure
\begin{equation}
	\begin{tikzpicture}[node distance=0.5 and 0.5]
		\node (s0) {$\ket{\Psi}\equiv\begin{pmatrix}
	\alpha_0 & \alpha_1
\end{pmatrix}^\top\equiv$};
		\node[vertex,right=of s0]  (dd0) {};
		\node[terminal, below=of dd0] (t0) {};
		\draw[edge={1}{0}] ($(dd0)+(0,0.5cm)$) -- (dd0);
		\draw[edge0={0.707}{0}] (dd0) to node[midway, left] {$\alpha_0$} (t0);
		\draw[edge1={0.707}{0}] (dd0) to node[midway, right] {$\alpha_1$} (t0);
	\end{tikzpicture}.
\end{equation}
It consists of a single \emph{node} with one \emph{incoming edge} that represents the entry point into the decision diagram as well as two \emph{successors} that represent the splitting shown from \autoref{eq:splitting} and end in a \emph{terminal} node (the black box).
The state's amplitudes are annotated to the respective edges.
Edges without annotations correspond to an edge weight of 1.

\begin{example}
Consider the computational basis states $\ket{0}$ and $\ket{1}$.
Then, the corresponding decision diagrams have the structures
\begin{equation}
	\begin{tikzpicture}[node distance=0.5 and 0.5]
		\node (s0) {$\ket{0}\equiv\begin{pmatrix}
	1 & 0
\end{pmatrix}^\top\equiv$};
		\node[vertex,right=of s0] (dd0) {};
		\node[terminal, below=of dd0] (t0) {};
		\draw[edge={1}{0}] ($(dd0)+(0,0.5cm)$) -- (dd0);
		\draw[edge0={1}{0}] (dd0) to node[midway,left] {$1$} (t0);
		\draw[edge1={1}{0}] (dd0) to ++(-50:0.35) node[zerostub] {};	
		
		\node[right=of dd0] (s1) {$\ket{1}\equiv\begin{pmatrix}
	0 & 1
\end{pmatrix}^\top\equiv$};
		\node[vertex,right=of s1] (dd1) {};
		\node[terminal, below=of dd1] (t1) {};
		\draw[edge={1}{0}] ($(dd1)+(0,0.5cm)$) -- (dd1);
		\draw[edge1={1}{0}] (dd1) to node[midway,right] {$1$} (t1);
		\draw[edge0={1}{0}] (dd1) to ++(-130:0.35) node[zerostub] {};	
	\end{tikzpicture} .
\end{equation}
In each of the cases, one of the successors ends in the terminal node, while the other ends in a \emph{zero stub} (indicated by a black dot)---notably resembling the corresponding vector descriptions.
\end{example}

Building off the intuition from a single-site state, we can move to larger systems.

\begin{example}\label{ex:multiqubitdd}
	Consider the following statevector of a three-site system:
	\begin{equation}
		\ket{\Psi} = \begin{pmatrix} \frac{1}{2\sqrt{2}} & \frac{1}{2\sqrt{2}} &  \frac{1}{2} & 0 & \frac{1}{2\sqrt{2}} & \frac{1}{2\sqrt{2}} & \frac{1}{2} & 0\end{pmatrix}^T
	\end{equation}
	Then, $\ket{\Psi}$ can be recursively split into equally-sized parts similar to \autoref{eq:splitting}, i.e.,
	\begin{equation}
                \overbrace{
                \overbrace{\begin{matrix}
                \overbrace{\begin{matrix}
                \bigl( \overset{\ket{000}}{
                \begin{matrix} \frac{1}{2\sqrt{2}}
                \end{matrix} }
                & \overset{\ket{001}}{
                \begin{matrix} \frac{1}{2\sqrt{2}}
                \end{matrix} }
                \end{matrix}}^{\ket{00q_0}}
                & \overbrace{
                \begin{matrix}\overset{\ket{010}}{
                \begin{matrix} \frac{1}{2}
                \end{matrix}}
                & \overset{\ket{011}}{
                \begin{matrix} 0
                \end{matrix} }
                \end{matrix}}^{\ket{01q_0}}
                \end{matrix}}^{\ket{0q_1q_0}}
                \ \
                \overbrace{\begin{matrix}
                \overbrace{\begin{matrix}
                \overset{\ket{100}}{
                \begin{matrix} \frac{1}{2\sqrt{2}}
                \end{matrix} }
                & \overset{\ket{101}}{
                \begin{matrix} \frac{1}{2\sqrt{2}}
                \end{matrix} }
                \end{matrix}}^{\ket{10q_0}}
                & \overbrace{
                \begin{matrix} \overset{\ket{110}}{
                \begin{matrix} \frac{1}{2}
                \end{matrix} }
                & \overset{\ket{111}}{
                \begin{matrix} 0
                \end{matrix} } \bigr)^\top
                \end{matrix}}^{\ket{11q_0}}
                \end{matrix}}^{\ket{1q_1q_0}}
                }^{\ket{q_2q_1q_0}} \\\\\\\\\ ,
    \end{equation}
    where $q_2, q_1, q_0 \in \{0, 1\}$. 
 	This directly translates to the decision diagram formalism:
 	\begin{equation}
 		\begin{tikzpicture}[node distance=0.5 and 1.0]
 		\node[vertex] (q2) {$q_2$};
 		\node[vertex,below left=0.5 and 1.75 of q2] (q1a) {$q_1$};
 		\node[vertex,below right=0.5 and 1.75 of q2] (q1b) {$q_1$};
 		\node[vertex,below left= of q1a] (q0a) {$q_0$};
 		\node[vertex,below right= of q1a] (q0b) {$q_0$};
 		\node[vertex,below left= of q1b] (q0c) {$q_0$};
 		\node[vertex,below right= of q1b] (q0d) {$q_0$};
 		\node[terminal, below= of q0a] (ta) {};
 		\node[terminal, below= of q0b] (tb) {};
 		\node[terminal, below= of q0c] (tc) {};
 		\node[terminal, below= of q0d] (td) {};
 	
		\draw[edge={1}{0}] ($(q2)+(0,0.5cm)$) -- (q2);
		
		\draw[edge0={1}{0}] (q2) to (q1a);		      
            \draw[edge1={1}{0}] (q2) to (q1b);
		
		\draw[edge0={1}{0}] (q1a) to (q0a);
		\draw[edge1={1}{0}] (q1a) to (q0b);
		
		\draw[edge0={1}{0}] (q1b) to (q0c);			
		\draw[edge1={1}{0}] (q1b) to (q0d);
		
		\draw[edge0={1}{0}] (q0a) to node[midway, left] {$\frac{1}{2\sqrt{2}}$} (ta);
  		\draw[edge1={1}{0}] (q0a) to node[midway, right] {$\frac{1}{2\sqrt{2}}$} (ta);

  		\draw[edge0={1}{0}] (q0b) to node[midway,left] {$\frac{1}{2}$} (tb);
      	\draw[edge1={1}{0}] (q0b) to ++(-50:0.35)  node[zerostub] {};

        \draw[edge0={1}{0}] (q0c) to node[midway, left] {$\frac{1}{2\sqrt{2}}$} (tc);
      	\draw[edge1={1}{0}] (q0c) to node[midway, right] {$\frac{1}{2\sqrt{2}}$} (tc);
       
    	\draw[edge0={1}{0}] (q0d) to node[midway,left] {$\frac{1}{2}$} (td);
        \draw[edge1={1}{0}] (q0d) to ++(-50:0.35) node[zerostub] {};
	\end{tikzpicture}
 	\end{equation}
 	Each level of the decision diagram consists of decision nodes with corresponding left and right successor edges. These successors represent the path that leads to an amplitude where the local quantum system (corresponding to the \emph{level} of the node, annotated here with the labels) is in the $\ket{0}$ (left successor) or the $\ket{1}$ state (right successor).
\end{example}
 	
At this point, this has just been a one-to-one translation between the statevector and a fancy graphical representation.
The core, unique feature of decision diagrams is that their graph structure allows redundant parts to be merged in the representation instead of representing them repeatedly.

\begin{example}\label{ex:reduction}
	Observe how, in the previous example, the left and the right successor of the top-level ($q_0$) node lead to exactly the same structure.
	As a result, the whole sub-diagram does not need to be represented twice, i.e.,
	\begin{equation}
		\begin{tikzpicture}[node distance=0.5 and 0.75]
 		\node[vertex] (q2) {$q_2$};
 		\node[vertex,below= of q2] (q1) {$q_1$};
 		\node[vertex,below left= of q1] (q0a) {$q_0$};
 		\node[vertex,below right= of q1] (q0b) {$q_0$};
 		\node[terminal, below= of q0a] (ta) {};
 		\node[terminal, below= of q0b] (tb) {};
 	
		\draw[edge={1}{0}] ($(q2)+(0,0.5cm)$) -- (q2);
		
		\draw[edge0={1}{0}] (q2) to (q1);
		\draw[edge1={1}{0}] (q2) to (q1);
		
		\draw[edge0={1}{0}] (q1) to (q0a);
		\draw[edge1={1}{0}] (q1) to (q0b);
		
		\draw[edge0={1}{0}] (q0a) to node[midway, left] {$\frac{1}{2\sqrt{2}}$} (ta);
  		\draw[edge1={1}{0}] (q0a) to node[midway, right] {$\frac{1}{2\sqrt{2}}$} (ta);

  		\draw[edge0={1}{0}] (q0b) to node[midway,left] {$\frac{1}{2}$} (tb);
      	\draw[edge1={1}{0}] (q0b) to ++(-50:0.35)  node[zerostub] {};
	\end{tikzpicture}
	\end{equation}
	From a memory perspective, this reduction alone has compressed the overall memory required to represent the state by 50\%.
\end{example}

Identifying redundancies in these kind of representations heavily depends on employing what is referred to as a \emph{normalization scheme} for the decision diagrams nodes \cite{niemannQMDDsEfficientQuantum2016}.
Such a normalization scheme makes sure that two decision diagram nodes that represent the same functionality do indeed have the same numerical structure.
In computer science, this property is referred to as \emph{canonicity} \cite{niemannQMDDsEfficientQuantum2016}.

The most commonly used and practically relevant normalization scheme is to normalize the outgoing edges of a node by dividing both weights by the norm of the vector containing both edge weights and adjusting the incoming edges accordingly~\cite{hillmichJustRealThing2020}.
This normalizes the sum of the squared magnitudes of the outgoing edge weights to~1 and is consistent with the quantum semantics, where basis states \(\ket{0}\) and \(\ket{1}\) are observed after measurement with probabilities that are squared magnitudes of the respective weights.
Normalization is recursively applied in a bottom-up fashion to ensure that every possible redundancy is being taken into account.

\begin{example}\label{ex:reduced}
Considering the decision diagram from the previous example, this results in the following \emph{normalized} and \emph{reduced} decision diagram:
\begin{equation}
		\begin{tikzpicture}[node distance=0.55 and 0.55]
 		\node[vertex] (q2) {$q_2$};
 		\node[vertex,below= of q2] (q1) {$q_1$};
 		\node[vertex,below left= of q1] (q0a) {$q_0$};
 		\node[vertex,below right= of q1] (q0b) {$q_0$};
 		\node[terminal, below= of q0a] (ta) {};
 		\node[terminal, below= of q0b] (tb) {};
 	
		\draw[edge={1}{0}] ($(q2)+(0,0.5cm)$) -- (q2);
		
		\draw[edge0={1}{0}] (q2) to node[midway, left] {$\frac{1}{\sqrt{2}}$} (q1);
		\draw[edge1={1}{0}] (q2) to node[midway, right] {$\frac{1}{\sqrt{2}}$} (q1);
		
		\draw[edge0={1}{0}] (q1) to node[midway, above left] {$\frac{1}{\sqrt{2}}$} (q0a);
		\draw[edge1={1}{0}] (q1) to node[midway, above right] {$\frac{1}{\sqrt{2}}$} (q0b);
		
		\draw[edge0={1}{0}] (q0a) to node[midway, left] {$\frac{1}{\sqrt{2}}$} (ta);
  		\draw[edge1={1}{0}] (q0a) to node[midway, right] {$\frac{1}{\sqrt{2}}$} (ta);

  		\draw[edge0={1}{0}] (q0b) to node[midway,left] {$1$} (tb);
      	\draw[edge1={1}{0}] (q0b) to ++(-50:0.35)  node[zerostub] {};
	\end{tikzpicture}
	\end{equation}
 The first two levels ($q_2$ and $q_1$) of the above diagram naturally encode that the respective sites have a $50/50$ ($\vert1/\sqrt{2}\vert^2 = 0.5$) probability to be in $\ket{0}$ and $\ket{1}$. Meanwhile, the bottom level ($q_0$) encodes that the probability of $q_0$ depends on the state of $q_1$.
 If $q_1$ is in the $\ket{0}$ state (following the left successor), then $q_0$ has probability $0.5$ in both $\ket{0}$ or $\ket{1}$. If $q_1$ is in the $\ket{1}$ state (following the right successor), it is guaranteed that the remaining site is in the $\ket{0}$ state.


\end{example}

Overall, statevectors are represented as decision diagrams conceptionally equivalent to halving the vector in a recursive fashion until it is fully decomposed.
The key idea is to exploit redundancies in the resulting diagrams to create a more compact representation.
Some interesting properties that are worth pointing out:
\begin{itemize}
	\item Decision diagrams can be initialized in their compact form (as, e.g., shown in \autoref{ex:reduced}). There is no need to create the maximally large decision diagram (as, e.g., shown in \autoref{ex:multiqubitdd}) or work directly from the statevector representation at any point in a calculation.
	\item Determining a particular amplitude of the represented state corresponds to multiplying the edge weights along a single-path traversal from the top edge of the decision diagram (called its \emph{root}) to a terminal node.
	\item The efficiency of decision diagrams is commonly measured by their \emph{size}, i.e., the number of nodes in the decision diagram---the smaller the number of nodes, the higher the compaction achieved by the data structure. Note that the terminal (node) is typically not counted towards the size of a decision diagram. 
	\item Any product state naturally has a decision diagram consisting of a single node per site. Highly-redundant states such as the GHZ state or the W state also have decision diagrams whose size (i.e., the number of nodes) is linear in the number of sites. A compact DD does not, however, correlate to the state being trivial.
	\item DDs are no "silver bullet." The worst case size of decision diagrams, corresponding to states with no redundancy, is still exponential in the number of sites. More specifically, a maximally large decision diagram has \mbox{$1+2^1+2^2+\dots+2^{L-1} = 2^L-1$} nodes for $L$ sites.
    \item In implementation, redundancy in the complex edge weights is equivalent to comparison of floating point numbers within some tolerance.
	\item To reduce visual clutter in illustrations of decision diagrams, edge weights are commonly not annotated explicitly, but their magnitude and phase is reflected in the thickness and the color of the respective edge. 
	In addition, to make the correspondence of the individual levels in a decision diagram to a system's sites more explicit, the nodes are frequently annotated with the site's index as an identifier.
	See~\cite{willeVisualizingDecisionDiagrams2021} for further details on common techniques for the visualization of decision diagrams.
\end{itemize}

\subsection{Representing Quantum Operators}
Quantum operators are fundamentally described by (\mbox{complex-valued}) matrices.
Just like moving from \emph{Matrix Product State} (MPS) representations to \emph{Matrix Product Operator} (MPO) representations in the domain of tensor networks, matrix decision diagrams are a natural extension of vector decision diagrams by an additional dimension.
To this end, consider the base case of a $2\times 2$ matrix $U$, i.e., 
\begin{align}
	U &= \begin{pmatrix}
		U_{00} & U_{01} \\ U_{10} & U_{11}
	\end{pmatrix} \notag\\ &= U_{00} \ket{0}\!\bra{0} + U_{01} \ket{1}\!\bra{0} + U_{10} \ket{0}\!\bra{1} + U_{11} \ket{1}\!\bra{1} .
\end{align}
Then, the decision diagram representing this matrix has the  structure
\begin{equation}
	\begin{tikzpicture}[node distance=0.5 and 0.5]
		\node (s0) {$U\equiv$};
		\node[vertex,right=of s0] (n1) {};
		\node[terminal, below left= 0.75 and 2.1 of n1] (t0) {};
		\node[terminal, below left= 0.75 and 0.7 of n1] (t1) {};
		\node[terminal, below right= 0.75 and 0.7 of n1] (t2) {};
		\node[terminal, below right= 0.75 and 2.1 of n1] (t3) {};
		\draw[edge={1}{0}] ($(n1)+(0,0.5cm)$) -- (n1);
		
		\draw[medge={1}{0}{-130}] (n1) to node[below left] {$U_{00}$} (t0);
		\draw[medge={1}{0}{-100}] (n1) to node[below] {$U_{01}$} (t1);
		\draw[medge={1}{0}{-80}] (n1) to node[below] {$U_{10}$} (t2);
		\draw[medge={1}{0}{-50}] (n1) to node[below right] {$U_{11}$} (t3);
	\end{tikzpicture} \\\\\ ,
\end{equation}
which again resembles the general structure of the matrix.
Note that $U_{ij}$ can be interpreted as the transformation of $\ket{j}$ to $\ket{i}$.

\begin{example}
	The following shows decision diagram representations for selected \mbox{single-qubit} operations:
	\begin{equation}\label{eq:single-qubit}
        \begin{tikzpicture}[node distance=0.5 and 0.125]
        \node (id) {$
		I = \begin{pmatrix}
            1 & 0 \\
            0 & 1
        \end{pmatrix} \equiv$};
        \node[vertex,right=of id] (n1) {};
		\node[terminal, below left=of n1] (t0) {};
		\node[terminal, below right=of n1] (t1) {};
		\draw[edge={1}{0}] ($(n1)+(0,0.5cm)$) -- (n1);
		\draw[medge={1}{0}{-130}] (n1) to node[left] {$1$} (t0);
		\draw[medge={1}{0}{-100}] (n1) to ++(-100:0.35)  node[zerostub] {};
		\draw[medge={1}{0}{-80}] (n1) to ++(-80:0.35)  node[zerostub] {};
		\draw[medge={1}{0}{-50}] (n1) to node[right] {$1$} (t1);
		
		\node[right=1 of n1] (x) {$
		X = \begin{pmatrix}
            0 & 1 \\
            1 & 0
        \end{pmatrix} \equiv$};
        \node[vertex,right=of x] (x1) {};
		\node[terminal, below left=of x1] (tx0) {};
		\node[terminal, below right=of x1] (tx1) {};
		\draw[edge={1}{0}] ($(x1)+(0,0.5cm)$) -- (x1);
		\draw[medge={1}{0}{-130}] (x1) to ++(-130:0.35)  node[zerostub] {};
		\draw[medge={1}{0}{-100}] (x1) to node[left] {$1$} (tx0);
		\draw[medge={1}{0}{-80}] (x1) to node[right] {$1$} (tx1);
		\draw[medge={1}{0}{-50}] (x1) to ++(-50:0.35)  node[zerostub] {};
		
		\node[below=1 of id] (rz) {$R_{z}(\theta) = \begin{pmatrix}
        e^{-i\frac{\theta}{2}} & 0 \\
        0 & e^{i\frac{\theta}{2}}
        \end{pmatrix}\equiv$};
        \node[vertex,right=0.5 of rz] (rz1) {};
		\node[terminal, below left=of rz1] (tz0) {};
		\node[terminal, below right=of rz1] (tz1) {};
		\draw[edge={1}{0}] ($(rz1)+(0,0.5cm)$) -- (rz1);
		\draw[medge={1}{0}{-130}] (rz1) to node[left] {$e^{-i\frac{\theta}{2}}$} (tz0);
		\draw[medge={1}{0}{-100}] (rz1) to ++(-100:0.35)  node[zerostub] {};
		\draw[medge={1}{0}{-80}] (rz1) to ++(-80:0.35)  node[zerostub] {};
		\draw[medge={1}{0}{-50}] (rz1) to node[right] {$e^{i\frac{\theta}{2}}$} (tz1);
		
		\node[right= 0.5 of rz1] (eq) {$\equiv$};
        \node[vertex,right=0.5 of eq] (rz2) {};
		\node[terminal, below left=of rz2] (tz2) {};
		\node[terminal, below right=of rz2] (tz3) {};
		\draw[edge={1}{0}] ($(rz2)+(0,0.5cm)$) node[right] {$e^{-i\frac{\theta}{2}}$} -- (rz2);
		\draw[medge={1}{0}{-130}] (rz2) to node[left] {$1$} (tz2);
		\draw[medge={1}{0}{-100}] (rz2) to ++(-100:0.35)  node[zerostub] {};
		\draw[medge={1}{0}{-80}] (rz2) to ++(-80:0.35)  node[zerostub] {};
		\draw[medge={1}{0}{-50}] (rz2) to node[right] {$e^{i\theta}$} (tz3);
    \end{tikzpicture}
	\end{equation}
	The last equivalence demonstrates how a common factor between the edge weights can be pulled out and attached to the incoming (root) edge.
\end{example}

The generalization to larger matrices works analogously to the vector case.
To construct the decision diagram representing a matrix, the matrix is recursively split into quarters and the four elements correspond to the four successors of the node for representing that split.
As with vector decision diagrams, a normalization scheme is applied to ensure that the resulting data structure is canonical and redundancy can be exploited.
The conventional approach is to normalize all edge weights by the weight with the highest magnitude, selecting the leftmost one if multiple weights have the same magnitude.
It is important to note that this ensures that all complex numbers within the decision diagram have a magnitude of at most $1$, although this is subject to the implementation.

\begin{example}
Consider the maximally-entangling two-qubit $R_{xx}$ rotation represented by the matrix
        \begin{equation}
        R_{xx} \Bigl(\theta = \frac{\pi}{2} \Bigl) = \frac{1}{\sqrt{2}}\begin{pmatrix}
            1 & 0 & 0 & -i \\
            0 & 1 & -i & 0 \\
            0 & -i & 1 & 0 \\
            -i & 0 & 0 & 1
        \end{pmatrix} .
    \end{equation}
    This matrix is equivalent to blocks of $2 \times 2$ matrices corresponding to the identity $I$ and the Pauli-$X$ matrix, i.e.,
    \begin{equation}\label{eq:rxxmat}
        R_{xx} \Bigl(\theta = \frac{\pi}{2} \Bigl) = \frac{1}{\sqrt{2}}\begin{pmatrix}
            I & -iX \\
            -iX & I
        \end{pmatrix} .
    \end{equation}
    The corresponding (already reduced) decision diagram has the following structure:
    \begin{equation}
    	\begin{tikzpicture}[node distance=0.5 and 0.125]
    	\node[vertex] (top) {};
    	\draw[edge={1}{0}] ($(top)+(0,0.5cm)$) node[right] {$\frac{1}{\sqrt{2}}$} -- (top);
    	\node[below left=0.75 and 0.75 of top, vertex] (id) {};
    	\node[below right=0.75 and 0.75 of top, vertex] (x) {};
    	\node[terminal,below left=of id] (t0) {};
    	\node[terminal,below right=of id] (t1) {};
    	\node[terminal,below left=of x] (tx0) {};
    	\node[terminal,below right=of x] (tx1) {};
    	
    	\draw[medge={1}{0}{-130}] (top) to node[above left] {$1$} (id);
		\draw[medge={1}{0}{-100}] (top) to node[below] {$-i$} (x);
		\draw[medge={1}{0}{-80}] (top) to node[above right] {$-i$} (x);
		\draw[medge={1}{0}{-50}] (top) to node[below left] {$1$} (id);
    	
    	\draw[medge={1}{0}{-130}] (id) to node[left] {$1$} (t0);
		\draw[medge={1}{0}{-100}] (id) to ++(-100:0.35)  node[zerostub] {};
		\draw[medge={1}{0}{-80}] (id) to ++(-80:0.35)  node[zerostub] {};
		\draw[medge={1}{0}{-50}] (id) to node[right] {$1$} (t1);
    	
    	\draw[medge={1}{0}{-130}] (x) to ++(-130:0.35)  node[zerostub] {};
		\draw[medge={1}{0}{-100}] (x) to node[left] {$1$} (tx0);
		\draw[medge={1}{0}{-80}] (x) to node[right] {$1$} (tx1);
		\draw[medge={1}{0}{-50}] (x) to ++(-50:0.35)  node[zerostub] {};
    	
    	\end{tikzpicture}
    \end{equation}
    Notice how the decision diagram naturally resembles the structure of the matrix.
    The nodes at the bottom represent the identity and the $X$ matrix (cf.~\autoref{eq:single-qubit}) while the node at the top encodes the redundancy of the \mbox{upper-left} and the \mbox{bottom-right} quadrant as well as the upper-right and the lower-left quadrant in \autoref{eq:rxxmat}.
    Similar to \autoref{ex:reduction}, exploiting redundancy has halved the overall memory requirement.
\end{example}

Again, some interesting properties to point out:
\begin{itemize}
	\item Just as in the vector case, it is always possible to work with the reduced form of matrix decision diagrams right away, i.e., without ever constructing the \mbox{exponentially-sized}, \mbox{maximally-large} diagram.
	\item A maximally-large matrix decision diagram for an $L$-site two-level system has $\sum_{i=1}^L 4^{i-1} = \frac{1}{3}(4^L -1)$ nodes, with each node having up to 4 scalar edge weights, determined by sparsity, i.e., the number of zero stubs.
	\item Decision diagrams are not limited to local interactions. Even long-range interactions between arbitrary sites typically emit compact representations as decision diagrams. For example, any two-site interaction between arbitrary sites can be represented as a decision diagram with at most $1+4(L-1)$ nodes---an exponential reduction.
	\item Decision diagrams are not limited to two-qubit interactions either. For example, controlled quantum gates with arbitrarily many controls (such as the multi-controlled Toffoli gate) give rise to decision diagrams with a linear number of nodes.
\end{itemize}

\subsection{Fundamental Operations on Decision Diagrams}

Merely defining means for compactly representing any kind of state or operator does not yet
allow one to perform Hamiltonian simulation.
It is crucial to also define efficient means to work with or manipulate the resulting representations.
In the following, we demonstrate how the most fundamental operations for Hamiltonian simulation can be carried out within the decision diagram formalism and how they scale.
We will mostly focus on how operations are realized on vectors, since the concepts extend from vectors to matrices in a straight-forward fashion.

The main concept throughout all of these schemes is to recursively break the respective operations down into sub-computations. This decomposition then naturally translates to the recursive decomposition of decision diagrams.
As such, operations generally scale with the number of nodes in the involved decision diagrams.

%
\subsection*{\textbf{Kronecker Product}}

The Kronecker product is necessary for creating product states as well as chaining together local operations.
For vectors, it can be expressed as
\begin{equation}
    \ket{\Psi} \otimes \ket{\Phi} = \begin{pmatrix}
                   \Psi_{0} \ket{\Phi} \\
                   \Psi_{1} \ket{\Phi}
                  \end{pmatrix}
                  = \begin{pmatrix}
                   \Psi_{0} \begin{pmatrix} \Phi_0 \\ \Phi_1 \end{pmatrix} \\
                   \Psi_{1} \begin{pmatrix} \Phi_0 \\ \Phi_1 \end{pmatrix}
                  \end{pmatrix}.
\end{equation}
In the decision diagram formalism, this is one of the simplest operations to perform and is done by simply replacing the terminal nodes of the first decision diagram with the root node of the second decision diagram.
In case of the above example, this has the following form:
	\begin{equation}
		\begin{tikzpicture}[node distance=0.5 and 0.125]
    	\node[vertex] (top0) {$\Psi$};
    	\node[terminal, below=of top0] (t0) {};
    	\draw[edge={1}{0}] ($(top0)+(0,0.5cm)$) to (top0);
    	\draw[edge0={1}{0}] (top0) to node[midway, left] {$\Psi_0$} (t0);
    	\draw[edge1={1}{0}] (top0) to node[midway, right] (psi1) {$\Psi_1$} (t0);
    	
    	\node[vertex,right=1.5 of top0] (top1) {$\Phi$};
    	\node[terminal, below=of top1] (t1) {};
    	\draw[edge={1}{0}] ($(top1)+(0,0.5cm)$) -- (top1);
    	\draw[edge0={1}{0}] (top1) to node[midway, left] (phi0) {$\Phi_0$} (t1);
    	\draw[edge1={1}{0}] (top1) to node[midway, right] {$\Phi_1$} (t1);
    	
    	\node[] at ($(top0)!0.5!(top1)$) {$\otimes$};
    	
    	\node[vertex,right=1.5 of top1] (top2) {$\Psi$};
    	\node[vertex,below=of top2] (middle) {$\Phi$};
    	\node[terminal, below=of middle] (t2) {};
    	\draw[edge={1}{0}] ($(top2)+(0,0.5cm)$) -- (top2);
    	\draw[edge0={1}{0}] (top2) to node[midway, left] {$\Psi_0$} (middle);
    	\draw[edge1={1}{0}] (top2) to node[midway, right] {$\Psi_1$} (middle);
    	\draw[edge0={1}{0}] (middle) to node[midway, left] {$\Phi_0$} (t2);
    	\draw[edge1={1}{0}] (middle) to node[midway, right] {$\Phi_1$} (t2);
    	
    	\node[] at ($(top1)!0.5!(top2)$) {$=$};
		\end{tikzpicture}
\end{equation}
As such, its complexity is linear in the number of nodes of the first decision diagram.

\subsection*{\textbf{Addition}}
Standard vector addition can be recursively broken down according to
\begin{equation}
    \begin{split}
        \ket{\Psi} + \ket{\Phi} & = \begin{pmatrix} \Psi_0 \\ \Psi_1 \end{pmatrix} + \begin{pmatrix} \Phi_0 \\ \Phi_1 \end{pmatrix} \\
        & = w \begin{pmatrix} \alpha_0  \\ \alpha_1 \end{pmatrix} + w' \begin{pmatrix} \alpha'_0 \\ \alpha'_1 \end{pmatrix} \\
        & = \begin{pmatrix} w \alpha_0 + w' \alpha'_0 \\ w \alpha_1 + w' \alpha'_1 \end{pmatrix},
    \end{split} 
\end{equation}
where $w$ and $w'$ are common factors of the terms in $\ket{\Psi}$ and $\ket{\Phi}$, respectively.

In the decision diagram formalism, this corresponds to a simultaneous traversal of both decision diagrams from their roots to the terminal (multiplying edge weights along the way until the individual amplitudes are reached) and back again (accumulating the results of the recursive computations).
More precisely,
\begin{equation}
	\begin{tikzpicture}[node distance=0.5 and 0.25]
    	\node[vertex] (top0) {};
    	\node[vertex, dashed, below left=of top0] (t0) {$\Psi_0$};
    	\node[vertex, dashed, below right=of top0] (t1) {$\Psi_1$};
    	\draw[edge={1}{0}] ($(top0)+(0,0.5cm)$) to node[midway,right] {$w$} (top0);
    	\draw[edge0={1}{0}] (top0) to node[midway,left] {$\alpha_0$} (t0);
    	\draw[edge1={1}{0}] (top0) to node[midway,right] {$\alpha_1$} (t1);
    	
    	\node[vertex,right=2 of top0] (top1) {};
    	\node[vertex, dashed, below left=of top1] (t2) {$\Phi_0$};
    	\node[vertex, dashed, below right=of top1] (t3) {$\Phi_1$};
    	\draw[edge={1}{0}] ($(top1)+(0,0.5cm)$) to node[midway,right] {$w'$} (top1);
    	\draw[edge0={1}{0}] (top1) to node[midway,left] {$\alpha'_0$} (t2);
    	\draw[edge1={1}{0}] (top1) to node[midway,right] {$\alpha'_1$} (t3);
    	
    	\node[] at ($(top0)!0.5!(top1)$) (plus) {$+$};
    	
    	\node[vertex,below=1.5 of plus] (top2) {};
    	\node[rectangle, draw, dashed, below left=of top2] (t4) {
    	\begin{tikzpicture}
    	\node[vertex, dashed] (a) {$\Psi_0$};
    	\draw[edge={1}{0},solid] ($(a)+(0,0.5cm)$) to node[midway,right] {$w\alpha_0$} (a);
    	\node[vertex, dashed,right=1.0 of a] (b) {$\Phi_0$};
    	\draw[edge={1}{0},solid] ($(b)+(0,0.5cm)$) to node[midway,right] {$w'\alpha'_0$} (b);
    	\node[right=0.5 of a] {$+$};
		\end{tikzpicture}};
    	\node[rectangle, draw, dashed, below right=of top2] (t5) {
    	\begin{tikzpicture}
    	\node[vertex, dashed] (a) {$\Psi_1$};
    	\draw[edge={1}{0},solid] ($(a)+(0,0.5cm)$) to node[midway,right] {$w\alpha_1$} (a);
    	\node[vertex, dashed,right=1.0 of a] (b) {$\Phi_1$};
    	\draw[edge={1}{0},solid] ($(b)+(0,0.5cm)$) to node[midway,right] {$w'\alpha'_1$} (b);
    	\node[right=0.5 of a] {$+$};
		\end{tikzpicture}};
    	\draw[edge={1}{0}] ($(top2)+(0,0.5cm)$) -- (top2);
    	\draw[edge0={1}{0}] (top2) to (t4);
    	\draw[edge1={1}{0}] (top2) to (t5); 
    	
    	\node[] at ($(plus)!0.6!(top2)$) {$=$};
		\end{tikzpicture} \\\\\ ,
\end{equation}
where the dashed nodes represent the respective successor decision diagrams.
Overall, this results in a complexity that is linear in the size of the larger decision diagram.

%

%
%
%

\subsection*{\textbf{Matrix-Vector Multiplication}}

Matrix-vector multiplication can be handled in a very similar fashion as addition.
Standard matrix-vector multiplication can be expressed as
\begin{equation}
    \begin{split}
    U\ket{\Psi} & = \begin{pmatrix} U_{00} & U_{01} \\
                                    U_{10} & U_{11} \end{pmatrix} \begin{pmatrix} \Psi_0 \\ \Psi_1 \end{pmatrix} \\
                & = w \begin{pmatrix} u_{00} & u_{01} \\
                u_{10} & u_{11} \end{pmatrix} w' \begin{pmatrix} \alpha_0 \\ \alpha_1 \end{pmatrix} \\
               & = ww' \begin{pmatrix} u_{00} \cdot \alpha_0 + u_{10} \cdot \alpha_1 \\
                u_{01} \cdot \alpha_0 + u_{11} \cdot \alpha_1 \end{pmatrix} .
    \end{split}
\end{equation}
This implies that a multiplication boils down to four smaller multiplications and two additions.
In the decision diagram formalism, this has the form
\begin{equation}\label{eq:mult}
	\begin{tikzpicture}[node distance=0.5 and 0.25]
    	\node[vertex] (top0) {};
    	\node[vertex, dashed, below left=0.5 and 1.5 of top0] (t0) {$U_{00}$};
    	\node[vertex, dashed, below left=0.5 and 0.75 of top0] (t1) {$U_{01}$};
    	\node[vertex, dashed, below right=0.5 and 0.75 of top0] (t2) {$U_{10}$};
    	\node[vertex, dashed, below right=0.5 and 1.5 of top0] (t3) {$U_{11}$};
    	
    	\draw[edge={1}{0}] ($(top0)+(0,0.5cm)$) to node[midway,right] {$w$} (top0);
    	\draw[medge={1}{0}{-130}] (top0) to node[midway,above left] {$u_{00}$} (t0);
    	\draw[medge={1}{0}{-100}] (top0) to node[midway,below] {$u_{01}$} (t1);
    	\draw[medge={1}{0}{-80}] (top0) to node[midway,below] {$u_{10}$} (t2);
    	\draw[medge={1}{0}{-50}] (top0) to node[midway,above right] {$u_{11}$} (t3);
    	
    	\node[vertex,right=3 of top0] (top1) {};
    	\node[vertex, dashed, below left=of top1] (t4) {$\Psi_0$};
    	\node[vertex, dashed, below right=of top1] (t5) {$\Psi_1$};
    	\draw[edge={1}{0}] ($(top1)+(0,0.5cm)$) to node[midway,right] {$w'$} (top1);
    	\draw[edge0={1}{0}] (top1) to node[midway,left] {$\alpha_0$} (t4);
    	\draw[edge1={1}{0}] (top1) to node[midway,right] {$\alpha_1$} (t5);
    	
    	\node[] at ($(top0)!0.65!(top1)$) (times) {$\bullet$};
    	
    	\node[vertex,below=1 of t2] (top2) {};
    	\node[rectangle, draw, dashed, below left=of top2] (t6) {
    	\begin{tikzpicture}
    	\node[vertex, dashed] (a) {$U_{00}$};
    	\draw[edge={1}{0},solid] ($(a)+(0,0.5cm)$) to node[midway,right] {$u_{00}$} (a);
    	\node[vertex, dashed,right=0.5 of a] (av) {$\Psi_0$};
		\draw[edge={1}{0},solid] ($(av)+(0,0.5cm)$) to node[midway,right] {$\alpha_{0}$} (av);
    	\node[vertex, dashed,right=0.5 of av] (b) {$U_{10}$};
    	\draw[edge={1}{0},solid] ($(b)+(0,0.5cm)$) to node[midway,right] {$u_{10}$} (b);
    	\node[vertex, dashed,right=0.5 of b] (bv) {$\Psi_1$};
    	\draw[edge={1}{0},solid] ($(bv)+(0,0.5cm)$) to node[midway,right] {$\alpha_{1}$} (bv);
    	\node[right=0.05 of av] {$+$};
    	\node[right=0.125 of a] {$\bullet$};
    	\node[right=0.125 of b] {$\bullet$};
		\end{tikzpicture}};
    	\node[rectangle, draw, dashed, below right=of top2] (t7) {
    	\begin{tikzpicture}
    	\node[vertex, dashed] (a) {$U_{01}$};
    	\draw[edge={1}{0},solid] ($(a)+(0,0.5cm)$) to node[midway,right] {$u_{01}$} (a);
    	\node[vertex, dashed,right=0.5 of a] (av) {$\Psi_0$};
		\draw[edge={1}{0},solid] ($(av)+(0,0.5cm)$) to node[midway,right] {$\alpha_{0}$} (av);
    	\node[vertex, dashed,right=0.5 of av] (b) {$U_{11}$};
    	\draw[edge={1}{0},solid] ($(b)+(0,0.5cm)$) to node[midway,right] {$u_{11}$} (b);
    	\node[vertex, dashed,right=0.5 of b] (bv) {$\Psi_1$};
    	\draw[edge={1}{0},solid] ($(bv)+(0,0.5cm)$) to node[midway,right] {$\alpha_{1}$} (bv);
    	\node[right=0.05 of av] {$+$};
    	\node[right=0.125 of a] {$\bullet$};
    	\node[right=0.125 of b] {$\bullet$};
		\end{tikzpicture}};
    	\draw[edge={1}{0}] ($(top2)+(0,0.5cm)$) to node[midway, right] {$ww'$} (top2);
    	\draw[edge0={1}{0}] (top2) to (t6);
    	\draw[edge1={1}{0}] (top2) to (t7); 
    	
    	\node[] at ($(t2)!0.5!(top2)$) {$=$};
		\end{tikzpicture} \\\\\\ ,
\end{equation}
where the dashed nodes again represent the respective successor decision diagrams.
Overall, this results in a complexity that scales with the product of the size of both decision diagrams.

\subsection*{\textbf{Inner Product}}
Computing the inner product of two vectors can be recursively broken down according to
\begin{equation}
    \begin{split}
        \braket{\Psi | \Phi} & = \begin{pmatrix} \Psi^*_0 & \Psi^*_1 \end{pmatrix} \begin{pmatrix} \Phi_0 \\ \Phi_1 \end{pmatrix} \\
        & = w^* \begin{pmatrix} \alpha^*_0 & \alpha^*_1 \end{pmatrix} w \begin{pmatrix} \alpha'_0 \\ \alpha'_1 \end{pmatrix} \\
        & = w^*w (\alpha^*_0 \alpha'_0 + \alpha^*_1 \alpha'_0)
    \end{split}
\end{equation}
This implies that the inner product boils down to two smaller inner product computations and adding the results. As with the matrix-vector multiplication, this is done recursively for each level of the decision diagram.
In the decision diagram formalism, this has the following form
\begin{equation}\label{eq:ip}
	\begin{tikzpicture}[node distance=0.5 and 0.25]
    	\node[vertex] (top0) {};
    	\node[vertex, dashed, below left=of top0] (t0) {$\Psi_0$};
    	\node[vertex, dashed, below right=of top0] (t1) {$\Psi_1$};
    	\draw[edge={1}{0}] ($(top0)+(0,0.5cm)$) to node[midway,right] {$w$} (top0);
    	\draw[edge0={1}{0}] (top0) to node[midway,left] {$\alpha_0$} (t0);
    	\draw[edge1={1}{0}] (top0) to node[midway,right] {$\alpha_1$} (t1);
    	
    	\node[vertex,right=2 of top0] (top1) {};
    	\node[vertex, dashed, below left=of top1] (t2) {$\Phi_0$};
    	\node[vertex, dashed, below right=of top1] (t3) {$\Phi_1$};
    	\draw[edge={1}{0}] ($(top1)+(0,0.5cm)$) to node[midway,right] {$w'$} (top1);
    	\draw[edge0={1}{0}] (top1) to node[midway,left] {$\alpha'_0$} (t2);
    	\draw[edge1={1}{0}] (top1) to node[midway,right] {$\alpha'_1$} (t3);
    	
    	\node[scale=3] at ($(top0)!0.5!(top1)$) (plus) {$\vert$};
    	\node[left=of top0,scale=3] (langle) {$\langle$};
    	\node[right=of top1,scale=3] (rangle) {$\rangle$};
    	
    	\node[rectangle, below left=1 and -2.5 of langle,left delimiter={(}, inner sep=0pt] (t4) {
    	\begin{tikzpicture}
    	\node[vertex, dashed] (a) {$\Psi_0$};
    	\draw[edge={1}{0},solid] ($(a)+(0,0.5cm)$) to node[midway,right] {$\alpha_0$} (a);
    	\node[vertex, dashed,right=0.5 of a] (b) {$\Phi_0$};
    	\draw[edge={1}{0},solid] ($(b)+(0,0.5cm)$) to node[midway,right] {$\alpha'_0$} (b);
    	\node[right=0.2 of a,scale=2] {$\vert$};
    	\node[left=0.1 of a,scale=2] {$\langle$};
    	\node[right=0.1 of b,scale=2] {$\rangle$};
		\end{tikzpicture}};
    	\node[rectangle,below right=1 and -2.5 of rangle,right delimiter={)}, inner sep=0pt] (t5) {
    	\begin{tikzpicture}
    	\node[vertex, dashed] (a) {$\Psi_1$};
    	\draw[edge={1}{0},solid] ($(a)+(0,0.5cm)$) to node[midway,right] {$\alpha_1$} (a);
    	\node[vertex, dashed,right=0.5 of a] (b) {$\Phi_1$};
    	\draw[edge={1}{0},solid] ($(b)+(0,0.5cm)$) to node[midway,right] {$\alpha'_1$} (b);
    	\node[right=0.2 of a,scale=2] {$\vert$};
    	\node[left=0.1 of a,scale=2] {$\langle$};
    	\node[right=0.1 of b,scale=2] {$\rangle$};
		\end{tikzpicture}};
    	
    	\node[] at ($(plus)!0.6!(top2)$) {$=$};
    	\node[] at ($(t4)!0.5!(t5)$) {$+$};
    	\node[left=of t4] {$w^*w'$};
		\end{tikzpicture} \ \ \ \ \ ,
\end{equation}
Overall, this results in a complexity that, just as addition, scales linearly with the size of the larger decision diagram.

\subsection*{\textbf{Expectation Value}}
Computing the expectation value of some observable $O$ for a given state $\ket{\Psi}$ can be reduced to a matrix-vector multiplication and an inner product computation as follows:
\begin{equation}
\braket{\Psi | O | \Psi} = \bra{\Psi} \Bigl( O \ket{\Psi} \Bigr) = \braket{\Psi | \tilde{\Psi}}
\end{equation}
This directly translates to decision diagrams via~\autoref{eq:mult} and~\autoref{eq:ip}.
The resulting computation has an overall complexity that scales with the product of the size of both decision diagrams.


%
%
%

\section{Experimental Evaluations} \label{sec:Results}
Using the concepts presented above, we implemented the first Hamiltonian simulation approach based on decision diagrams. To this end, we used the open-source DDSIM available at \url{https://github.com/cda-tum/mqt-ddsim} (which is part of the \emph{Munich Quantum Toolkit}, MQT) and extended it to support all necessary gates (such as the various \mbox{two-qubit} rotation gates) and operations (such as the expectation value).
Afterwards, we conducted several series of evaluations and comparisons to get insights about the performance of the resulting DD-based Hamiltonian simulation. More precisely,  we first analyzed different scaling characteristics of decision diagrams themselves followed by comparing their performance to other state-of-the-art techniques and considering selected best case scenarios. In this section, we summarize the obtained results and findings.


\subsection{Application of DDs to Hamiltonian Simulation Circuits}
\begin{figure}[t]
	\centering
	\includegraphics[width=\linewidth]{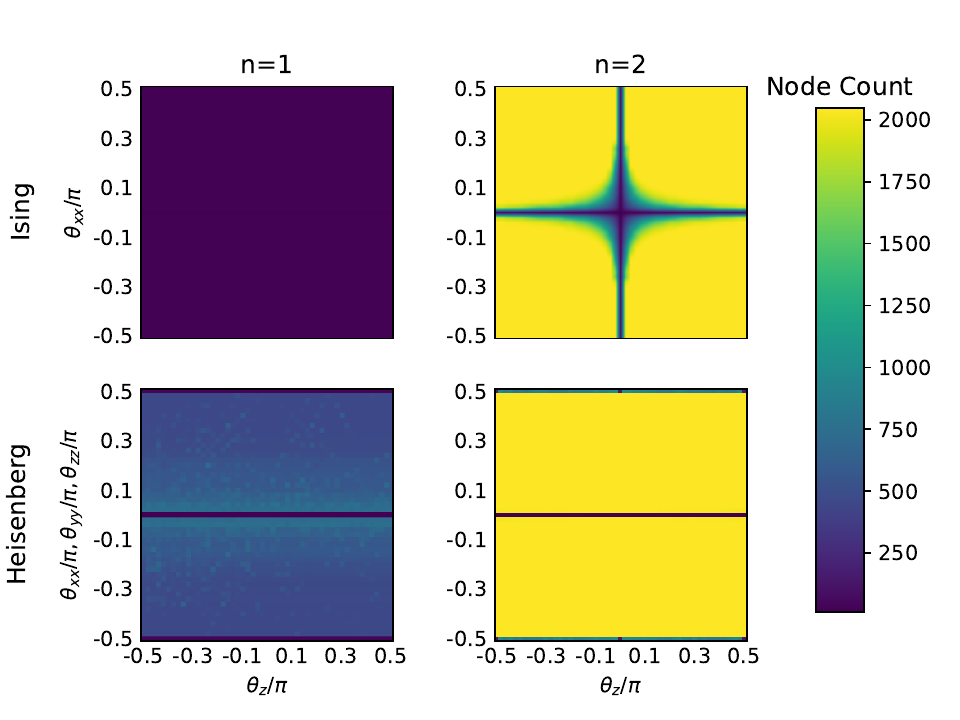}
	\vspace*{-0.6cm}

	\caption{Redundancy landscapes for the Ising and Heisenberg model with $L=12$ for $n=1$ and $n=2$ Trotter step applications of the selected angles.}
	\label{fig:RedundancyLandscapes}
	\vspace*{-0.3cm}

\end{figure}

In a first series of evaluations, we first analyzed the behavior of DDs in simulating various combinations of rotation angles in the Hamiltonian simulation circuits for the Ising and Heisenberg models described in \autoref{sec:HamiltonianSimulationBackground}. The results are shown in \autoref{fig:RedundancyLandscapes}. More precisely, we generated heatmaps, which we call \emph{redundancy landscapes}, that show the number of nodes in the DD after applying each circuit. This was performed for $n=1$ and $n=2$ Trotter steps on a system size $L=12$ initialized in the $\ket{0 \dots 0}$ state. The x- and y-axis correspond to the angle of the single- and two-site rotations, respectively. The node count scaling is normalized across both graphs---from maximal compaction (dark blue; linear regime) to almost no compaction (yellow; exponential regime).

These landscapes provide an abstract understanding of how each configuration of angles affects the size of the decision diagram. 
By using generalized angles based on the parameters of the Ising and Heisenberg Hamiltonians, these landscapes can also help identify the model parameters and timestep sizes that lead to compact decision diagrams.

From the results, it can be seen that the Ising model stays compact for one Trotter step, regardless of the angle combinations---achieving almost maximal compaction.
For two Trotter steps, the landscape begins to saturate such that there is almost no compaction for large angles.
However, for small angles, the size remains moderate even for multiple Trotter steps.

For the Heisenberg model, a single Trotter step causes larger, but still not maximally large, decision diagrams compared to the Ising model.
However, multiple Trotter steps causes the node count of the decision diagram to quickly grow with respect to the magnitude of the angles.

This suggests that DDs are likely to remain compact for Ising models and their derivatives, but require more work for simulating the Heisenberg model.

%
%

\subsection{Node Count based on System Size and Trotter Number}
\begin{figure}[t]
	\centering
	\includegraphics[width=\linewidth]{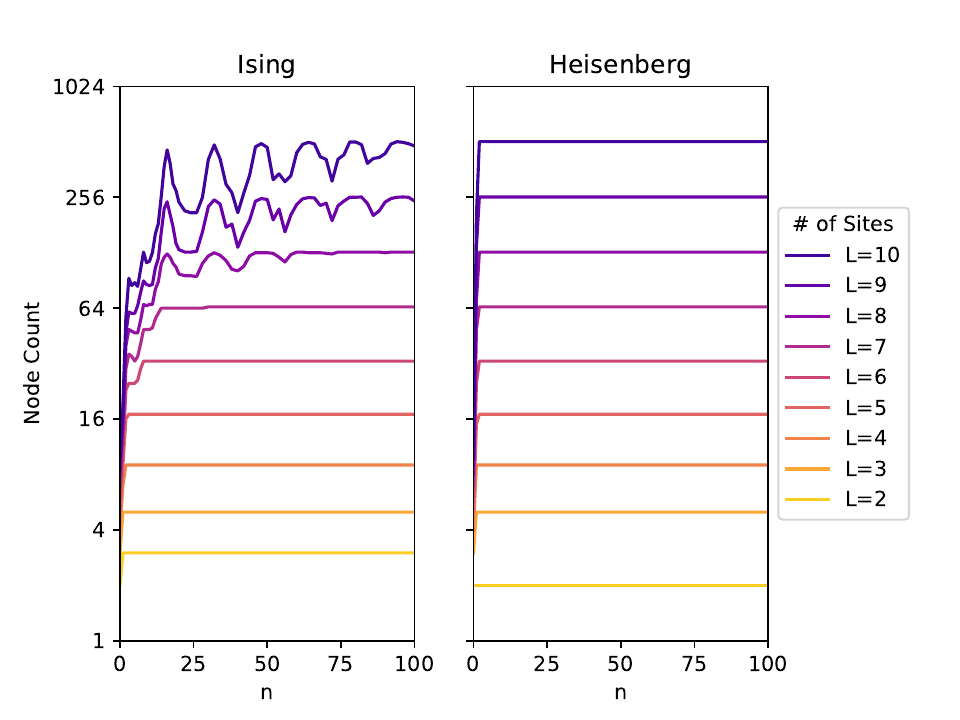}
	\vspace*{-0.6cm}

	\caption{Growth of node count in the decision diagrams for the Ising model ($J=1$, $g=0.001$) and Heisenberg model ($J_x=J_y=J_z=1, h=1)$ with timestep size $\delta t = 0.1$.}
	\label{fig:NodeScaling}
	\vspace*{-0.3cm}
\end{figure}

In a second series of evaluations, we evaluated the node count in the DD as various number of Trotter steps are applied to several system sizes. The results according to the Ising and Heisenberg models are depicted in \autoref{fig:NodeScaling}. In this figure, multiple Trotter steps are applied to systems ranging from size $L=2$ to $L=10$.
The circuits' angles are fixed and were selected based on the redundancy landscapes seen in the previous section. The Ising model shown has a normalized interaction parameter of $J=1$ and weak transverse field of $g=0.001$. The Heisenberg model has interaction parameters of $J_x=J_y=J_z=1$ and a field of $h=1$. The timestep size in both cases is $\delta t = 0.1$.

Several insights can be drawn from the plotted results. 
The initial slope for the first few Trotter steps has the greatest impact on scaling, with a maximum node count of $2^{L-1}$ in both models. 

\begin{figure*}[t]
	\centering
	\includegraphics[width=.8\linewidth]{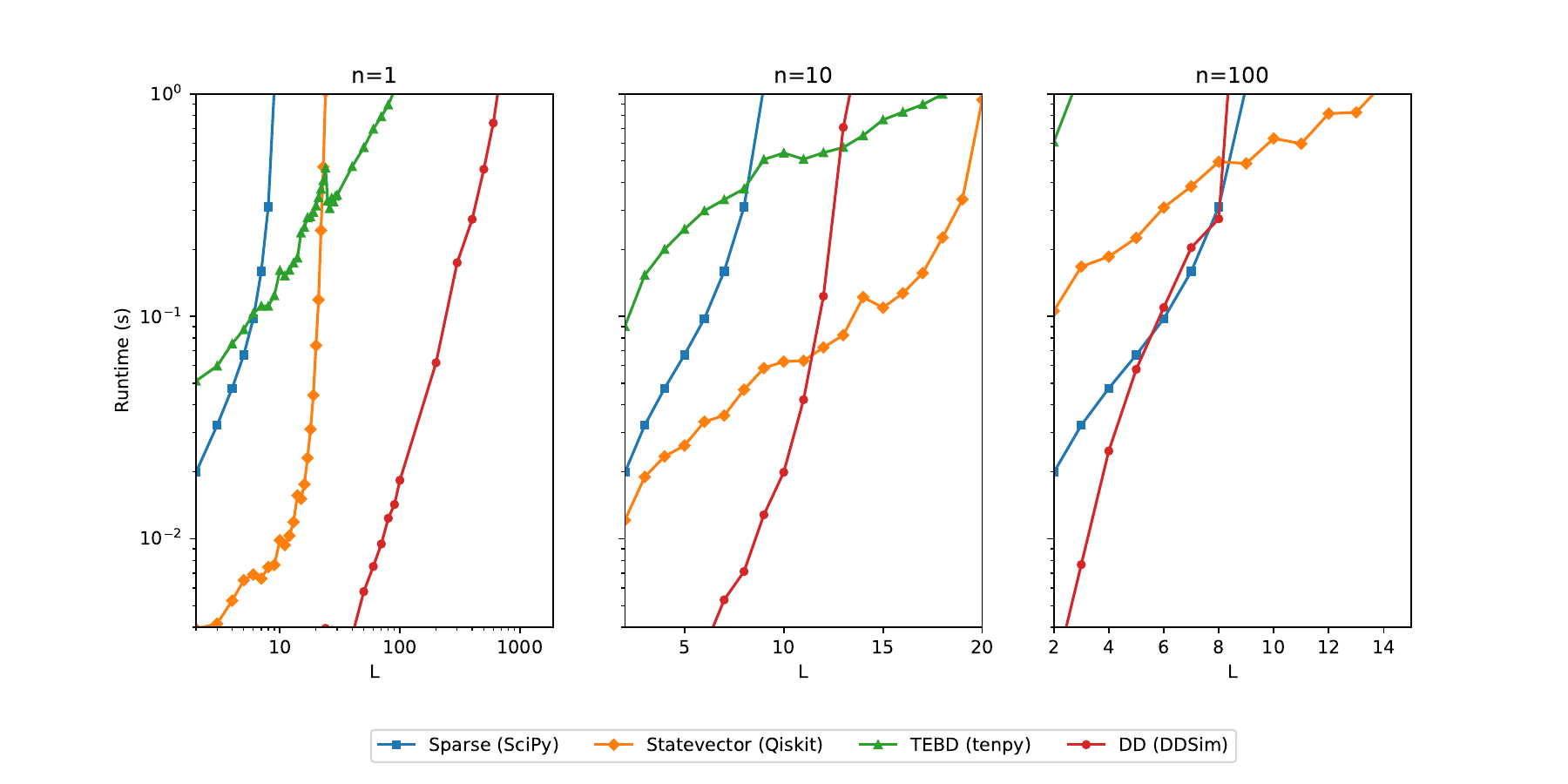}
	\vspace{-.5cm}
	\caption{Runtime scaling of various methods for simulating the Ising model $J=1$, $g=0.001$, and timestep size $\delta t = 0.1$. Each corresponds to this being performed with $n$ Trotter steps.}
	\vspace{-.5cm}
	\label{fig:MethodComparison}
\end{figure*}

\begin{figure}[t]
	\centering
	\begin{subfigure}{\columnwidth}
		\includegraphics[width=.8\linewidth]{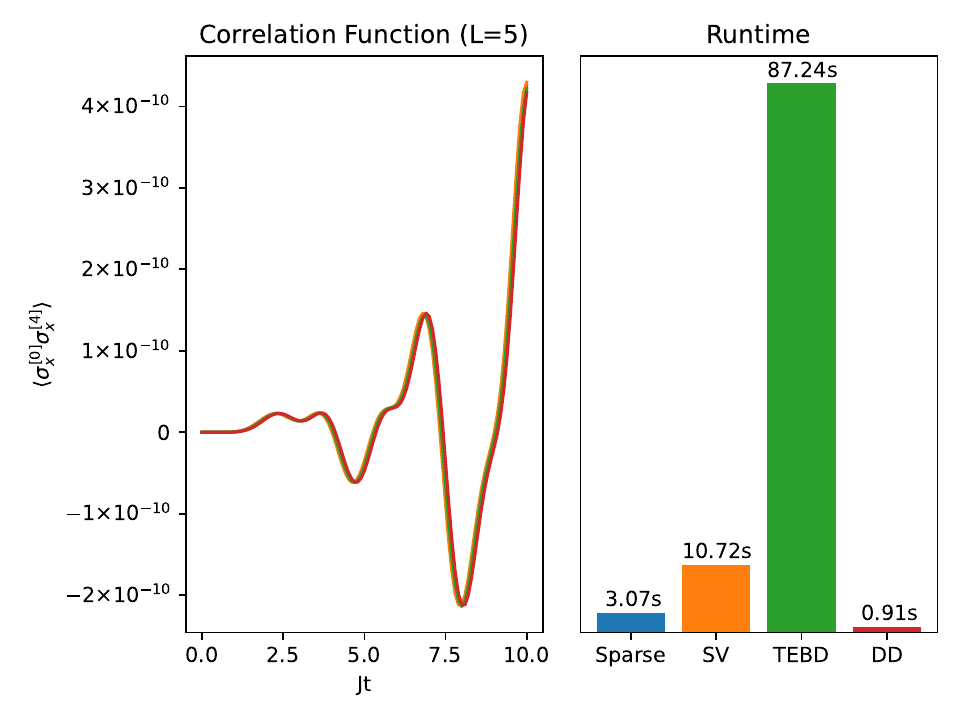}
		
		\caption{Time evolution of a two-site correlation function between the ends of a chain under the Ising model $J=1$, $g=0.001$ for $L=5$ sites}
		\label{fig:CorrelationExample}
	\end{subfigure}
	\begin{subfigure}{\columnwidth}
		\includegraphics[width=.8\linewidth]{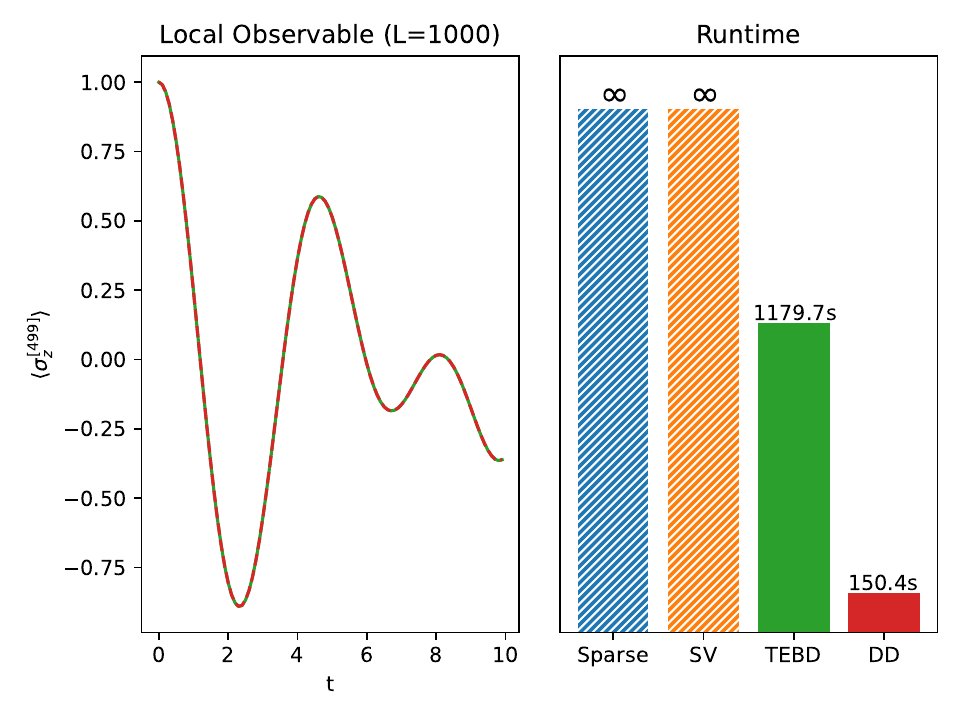}
		
		\caption{Time evolution of an Edwards-Anderson spin glass model with Gaussian distributed interaction parameters $J$ with mean 0 and standard deviation 1}
		\label{fig:SpinGlassExample}
	\end{subfigure}
	\vspace*{-0.6cm}
	\caption{Selected examples and their runtime requirement for different methods}
	\label{fig:Examples}
	\vspace*{-0.3cm}

\end{figure}

The Ising model exhibits an oscillation in node count, corresponding to times when the state is more redundant than others. 
This oscillation becomes larger as the system size grows, indicating that larger systems have more redundancies and more opportunities for significant compression.
The lower and upper bounds of this oscillation converge as the number of Trotter steps increases.

In contrast, the Heisenberg model converges to its maximum node count after only two Trotter steps---aligning with the growth shown in the redundancy landscapes.
At this timescale, the selected Heisenberg model parameters do not exhibit many redundancies in its DD representation.

These results suggest, again, that the Ising model can benefit from significant compression over long timescales---particularly for large systems. 
The oscillation in its node count also supports the need to engineer redundancy to maximize the benefits of the DDs. On the other hand, the Heisenberg model does not exhibit much redundancy.

%
%

\subsection{Comparison with Related Work} \label{sec:MethodComparison}
In a third series of evaluations, 
we compare the runtime scaling of the proposed DD-based Hamiltonian simulation 
against other simulation methods. More specifically, we compare against the time needed for an exact calculation with sparse methods implemented in SciPy \cite{2020SciPy-NMeth}, the Qiskit statevector simulator \cite{Qiskit}, and tenpy TEBD implementation \cite{tenpy}. These results are shown in \autoref{fig:MethodComparison}. Each plot shows the time for each method to perform the time evolution and calculation of the expectation value of a local observable $\langle \sigma_z \rangle$ at the center of the system. These results are averaged over 10 runs. This is done with the Ising model with paramters $J=1$, $g=0.001$, and timestep size $\delta t = 0.1$, but similar results are seen for other combinations such that this can be seen as a general guideline.

The results show that, for a single Trotter step, decision diagrams outperform both the Qiskit statevector simulator and tenpy TEBD, regardless of the system size. As the number of Trotter steps increases, however, the advantage of decision diagrams diminishes for larger systems. We notice that the runtime begins to converge with that of the sparse method. These findings suggest that decision diagrams are most advantageous for problems that can be solved in a single Trotter step. For multiple Trotter steps, they still provide a computational advantage for small systems, but the runtime converges, in the worst case, to the time required for a sparse method to compute the exact result.

These results show that DDs can indeed serve as a complementary alternative to current Hamiltonian simulation methods. As these results are based on the current state-of-the-art implementation of decision diagrams, it is expected that future work will lead to improved scaling for many Trotter steps and different models.

\subsection{Selected Examples}
In a final series of evaluations, we considered what we observed to be current best case scenarios for the considered DD-based Hamiltonian simulation, i.e.,
a 5-site Ising model and a 1000-site nearest-neighbor (Edwards-Anderson) spin glass chain without a field. 
These cases correspond to the results seen in \autoref{sec:MethodComparison}, i.e., small systems regardless of the number of Trotter steps and problems that can be solved in a single Trotter step, respectively.

The results of the 5-site Ising model can be seen in \autoref{fig:CorrelationExample}. We show the time evolution of the two-site correlation function between the two ends of a 5-site chain $\langle \sigma_x^{[0]} \sigma_x^{[4]} \rangle$, evolving under an Ising model with parameters $J=1$, $g=0.001$. This plot contains 100 sampled points, each Trotterized with a timestep $\delta t = 0.1$, i.e., $Jt=0.1$ requires 1 Trotter step and $Jt=10$ requires 100 Trotter steps. The runtime plot clearly shows that DDs outperform the methods typically used for a system of this size by an order of magnitude.

\autoref{fig:SpinGlassExample} shows the results of the spin glass model. This model has Ising interactions along one direction without a field. This is equivalent to the generalized Ising model presented in \autoref{eq:GeneralizedIsing} such that the parameters $J$ are chosen randomly from a Gaussian distribution with mean 0 and standard deviation~1. This means the terms of the Hamiltonian commute which eliminates the Trotter error in \autoref{eq:BCH}. The time evolution of a local observable $\langle \sigma_z^{[499]} \rangle$ at the center of the chain is plotted with 100 sampled points for timestep $\delta t = 0.1$. Each point is calculated with a single Trotter step. 

For systems of this size, tensor networks are currently the state-of-the-art method, as sparse methods and statevector simulators grow too large to be used efficiently. In this example, we also see that DDs outperform tensor networks by an order of magnitude.

These results indicate that the current implementation of DDs offers a significant computational advantage over other methods for small systems as well as for problems that can be solved in one Trotter step.

%
%
%

\section{Conclusion} \label{sec:Conclusion}
In this work, we proposed \emph{Decision Diagrams} (DDs) as a promising new data structure for Hamiltonian simulation. The obtained results show that DDs can efficiently handle highly redundant models, such as Ising-type models, surpassing statevector simulators and tensor networks in memory and runtime requirements. For problems solvable in a single Trotter step, DDs can efficiently scale to large systems, while for multiple Trotter steps, DDs show an advantage over other methods for small systems. As the number of Trotter steps increases, DDs eventually converge to the runtime requirements of sparse methods. However, our analysis of more complex models, such as the Heisenberg model, suggests that these models do not show as promising results, likely due to the lower level of redundancy. Further research is needed to explore DDs' potential for more complex models.

Still, these initial results already suggest that DDs could hold promise for problems that can be formulated as Ising models, such as QUBOs and graph problems. Additionally, preliminary application of DDs to long-range and higher-dimensional models have shown promise, although they exhibit redundancies that are not captured in the current decision diagram formalism. Therefore, future research could explore these higher-dimensional models and find ways to engineer redundancy in problem formulations in order to guarantee compact DDs.

We also acknowledge that the limitations of DDs are fundamentally different from those of current state-of-the-art methods, requiring a shift in perspective to fully utilize their benefits. Using DDs for Hamiltonian simulation opens up the possibility of implementing techniques from the field of graph theory, potentially allowing for sophisticated approximation techniques that could improve the scaling. The creation of approximation methods and finding ways to engineer redundancy in problem formulations are a promising next step for DDs and are left to future research.

In conclusion, this work demonstrates that DDs offer a promising new data structure for Hamiltonian simulation, especially for problems formulated with redundancies in mind. Future research can focus on refining the implementation of DDs and exploring their application to more complex models and problems. We anticipate that DDs will continue to play an important role in the development of efficient and accurate simulation methods.

\section*{Acknowledgments}
This work received funding from the European Research Council (ERC) under the European Union’s Horizon 2020 research and innovation program (grant agreement No. 101001318), was part of the Munich Quantum Valley, which is supported by the Bavarian state government with funds from the Hightech Agenda Bayern Plus, and has been supported by the BMWK on the basis of a decision by the German Bundestag through project QuaST.

\printbibliography

\end{document}